\documentclass[conference]{IEEEtran}
\IEEEoverridecommandlockouts
\usepackage{cite}
\usepackage{amsmath,amssymb,amsfonts}
\usepackage{algorithmic}
\usepackage{graphicx}
\usepackage{textcomp}
\usepackage{comment}
\usepackage{xcolor}
\usepackage[caption=false,font=footnotesize]{subfig}

\IEEEoverridecommandlockouts 
\IEEEaftertitletext{\vspace{-1\baselineskip}} 

\def\BibTeX{{\rm B\kern-.05em{\sc i\kern-.025em b}\kern-.08em
    T\kern-.1667em\lower.7ex\hbox{E}\kern-.125emX}}

\makeatletter
\newcommand{\linebreakand}{%
  \end{@IEEEauthorhalign}
  \hfill\mbox{}\par
  \mbox{}\hfill\begin{@IEEEauthorhalign}
}
\makeatother

\definecolor{lightblue}{HTML}{0f85bf} 

\begin{document}

\title{Self-Heating and Radiation Hardness Studies of 3nm GAA-FET-Based SRAM with Different Substrate Isolation Techniques}

\author{
\IEEEauthorblockN{Albert Lu}
\IEEEauthorblockA{
\textit{San Jose State University}\\
San Jose, California, USA \\
albert.lu@sjsu.edu \\ 
}
\and
\IEEEauthorblockN{Junipero Verbeke}
\IEEEauthorblockA{
\textit{San Jose State University}\\
San Jose, California, USA \\
junipero.verbeke@sjsu.edu \\ 
}
\and
\IEEEauthorblockN{Phil Oldiges}
\IEEEauthorblockA{\textit{Sandia National Laboratories} \\
New Mexico, USA \\ 
pjoldig@sandia.gov \\
}
\linebreakand
\IEEEauthorblockN{Reza Arghavani}
\IEEEauthorblockA{\textit{Sandia National Laboratories} \\
New Mexico, USA \\
rarghav@sandia.gov \\ 
}
\and
\IEEEauthorblockN{Hiu Yung Wong$^*$}
\IEEEauthorblockA{
\textit{San Jose State University}\\
San Jose, California, USA \\
hiuyung.wong@sjsu.edu}
}
\maketitle

\begin{abstract}
In this work, 3D full-domain 3 nm gate-all-around field-effect transistor (GAA-FET) static random access memories (SRAMs) with various substrate isolation techniques are simulated using Technology Computer-Aided Design (TCAD). In addition to the traditional bottom dielectric isolation (BDI), which isolates the source/drain (S/D) from the substrate (dubbed SD-BDI), and the punch-through stopper (PTS), a novel channel-BDI (C-BDI) is proposed, allowing S/D-to-substrate connection. The self-heating effect and radiation hardness due to various isolation techniques are studied. It is found that, firstly, the increase in self-heating due to BDI is negligible. Secondly, in the novel C-BDI, even without PTS, the increase in leakage current I$_{OFF}$ is minimal. Thirdly, for SD-BDI with underlap (to minimize stress relaxation), while I$_{OFF}$ increases, the static noise margin (SNM) remains unchanged and robust against single-event upset (SEU) even if the underlap is as much as 20 nm. Finally, all structures are immune to the alpha-particle SEU, and BDI enhances the radiation hardness substantially. Moreover, radiation hardness is insensitive to BDI thickness.
\end{abstract}

\vspace{1.5ex} 


\section{Introduction}

As semiconductor scaling progresses into the sub-5 nm regime, GAA-FETs have emerged as the leading architectural candidate to replace FinFETs \cite{b1}. By fully surrounding the channel with the gate, GAA-FETs offer superior electrostatic control, reduced short-channel effects, and improved drive current. Despite these advantages, the substrate subchannel is highly prone to parasitic leakage.

\begin{figure} 
    \centering
    \includegraphics[width=1\linewidth]{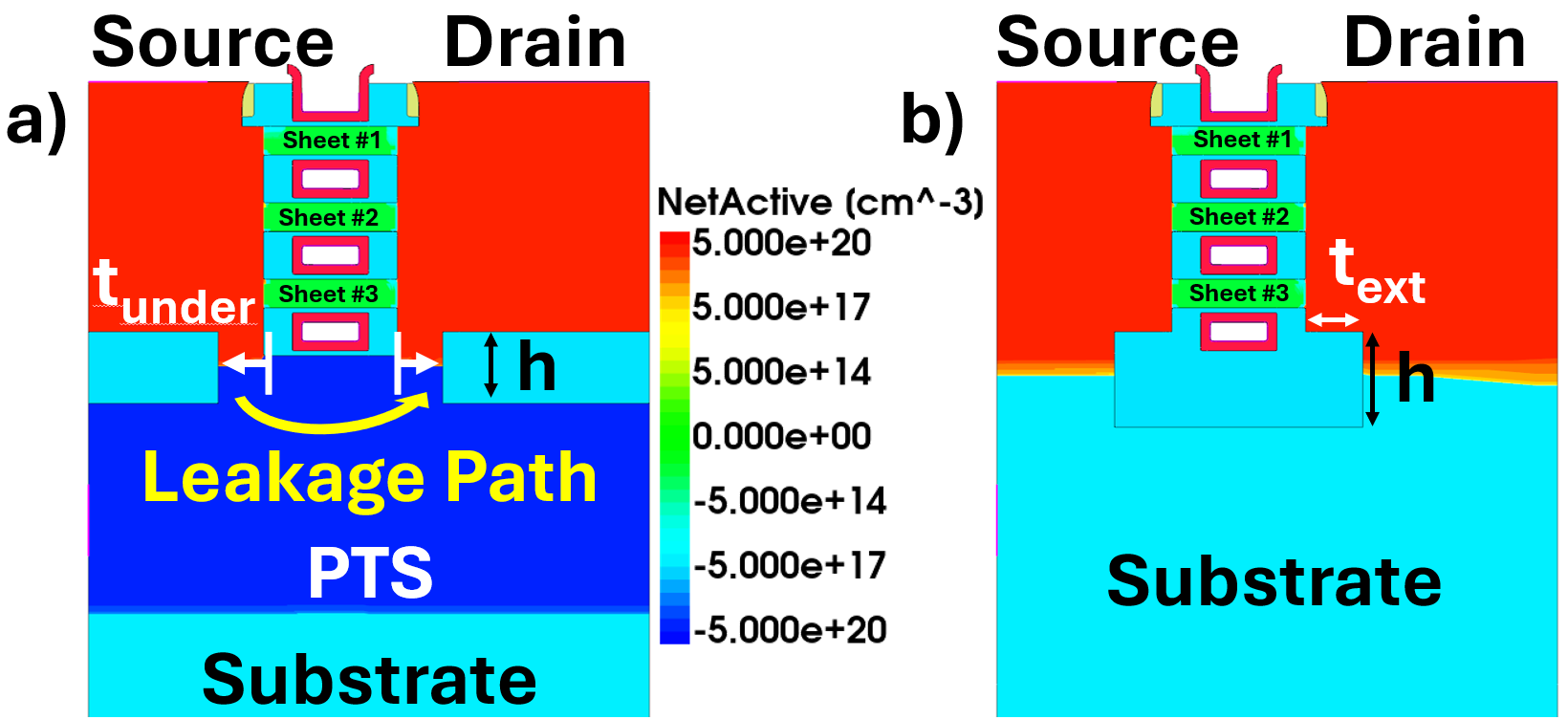}
    \caption{Cross-sections of the GAA-FETs under study. Both structures may or may not have PTS. $t_{under}$, $t_{ext}$, and $h$ are BDI parameters adjusted in this study. (a) GAA-FET with SD-BDI. The BDI is under the S/D. A Structure with PTS and SD-BDI underlap is shown. The substrate leakage path is indicated. (b) GAA with C-BDI, where BDI is under the channel. }
    \label{fig:GAACross-Section}
\end{figure}

To mitigate this subchannel leakage, several substrate isolation techniques have been studied in the literature, such as PTS and BDI \cite{b2}. PTS uses a relatively heavily doped layer beneath the channel and source/drain (S/D) to prevent substrate leakage. It is doped with the opposite type as the S/D. In the literature, BDI usually refers to an oxide layer used to isolate the S/D from the substrate and has two forms that are often studied \cite{b3, b6, b5}. ``Partial" BDI is often defined as when the oxide layer is placed only under the S/D (\textcolor{lightblue}{Fig.~\ref{fig:GAACross-Section}a}, where it also has a PTS). ``Full" BDI is when the oxide layer is placed under the S/D and channel as well (where the subchannel is at). 
PTS has the disadvantages of being more prone to process variations and higher parasitic capacitances. However, PTS allows a direct connection from the S/D to enable stress formation in the S/D and has the potential to reduce self-heating. BDI reduces subthreshold leakage and parasitic capacitance due to the oxide layer. BDI does not use doping, so it also allows for less consideration to process variations due to doping. Furthermore, BDI is also found to be beneficial for radiation hardness \cite{b6}. 

However, BDI has the disadvantage of severing the connection between the S/D and the substrate. This is suspected to worsen self-heating \cite{b7} due to the heat generated from the transistor not being able to escape as easily through the substrate. Self-heating can cause worse reliability, such as from bias temperature instability (BTI), hot carrier injection (HCI), and time-dependent dielectric breakdown (TDDB). Severing the connection from the S/D to the substrate can also cause stress relaxation. With BDI, the S/D epitaxy can only be grown well from the sheets, which reduces the quality of the S/D due to increased threading dislocations. Therefore, A novel channel-BDI (C-BDI) is proposed and studied in this work to alleviate the issue (\textcolor{lightblue}{Fig.~\ref{fig:GAACross-Section}b}). 


In this work, full-domain 3D TCAD simulations (which are necessary to capture layout effect \cite{b7x5}) are performed on 3 nm GAA-FET SRAMs with various substrate isolation techniques (PTS, SD-BDI, C-BDI) to study their effects on self-heating and radiation hardness.

\section{TCAD Simulations and Results}

\subsection{BDI Implementations and SRAM Design} 

\begin{table}[b]
\caption{3 nm Node Dimensions for GAA-FET.} 
\renewcommand{\arraystretch}{1.1}   
\label{tbl:3nm_node_dimensions}
\centering
\begin{tabular}{l c} 
\hline
\textbf{Parameter} & \textbf{Value} \\
\hline
Channel Doping  & No \\ 
\hline
Inner Spacer & 6 nm \\ 
\hline
Oxide Thickness (t$_{ox}$) & 1 nm \\
HfO$_2$ Thickness (t$_{hk}$) & 2 nm \\
\hline
Oxide Dielectric Constant & 3.5 \\
HfO$_2$ Dielectric Constant & 20 \\
\hline
Number of Sheets & 3 \\
Silicon Sheet Thickness (t$_{si}$) & 6 nm \\
Sheet-to-Sheet Spacing (t$_{sus}$) & 10 nm \\
Total Height & 48 nm \\
\hline
Gate Length (Gate Metal) (L$_G$) & 12 nm \\
Gate Length (Gate Metal + 2t$_{ox}$ + 2t$_{hk}$) (L$_{G1}$) & 18 nm \\
\hline
Sheet/Fin Width (NMOS) & 13 nm \\
Sheet/Fin Width (PMOS) & 17 nm \\
\hline
N-Type Work function & 4.25 eV \\
P-Type Work function & 4.85 eV\\
\hline
S/D Doping Outside of Inner Spacer & Yes \\
\hline
\end{tabular} 
\end{table}
 In this work, TCAD Sentaurus is used \cite{b8} (SDevice and SProcess) to study the characteristics of 3nm GAA-FET SRAMs of different substrate isolations and their radiation hardness. Philips Unified Mobility model, ballistic mobility model, Lombardi for surface scattering, impact ionization, SRH recombination, Auger recombination, and density gradient are turned on.

 GAA-FETs at the 3nm node are created based on the structure and dimensions from \cite{b1} and IRDS 2024 \cite{b10}. The dimensions used are shown in \textcolor{lightblue}{Table \ref{tbl:3nm_node_dimensions}}. Two different types of BDI are studied: SD-BDI and C-BDI (\textcolor{lightblue}{Fig. \ref{fig:GAACross-Section}}). SD-BDI places the BDI underneath the S/D, where an underlap value ($t_{under}$) and its height ($h$) can be adjusted. When $t_{under} = 0$, there is no electrical connection between the S/D and the substrate, and it is called a "partial BDI". A negative $t_{under}$ indicates that BDI extends under the channel, and if it fully covers the channel, it is a "full BDI". However, this physical discontinuity leads to S/D epitaxial stress relaxation and acts as a thermal barrier, leading to self-heating effects. In contrast, C-BDI places the BDI directly under the channel. It can be extended into the S/D regions (through $t_{ext}$). It should be noted that C-BDI can be created by combining and modifying the fabrication processes for full BDI and partial BDI.

    
    

\begin{figure}[t]  
    \centering
\includegraphics[width=0.8\linewidth]{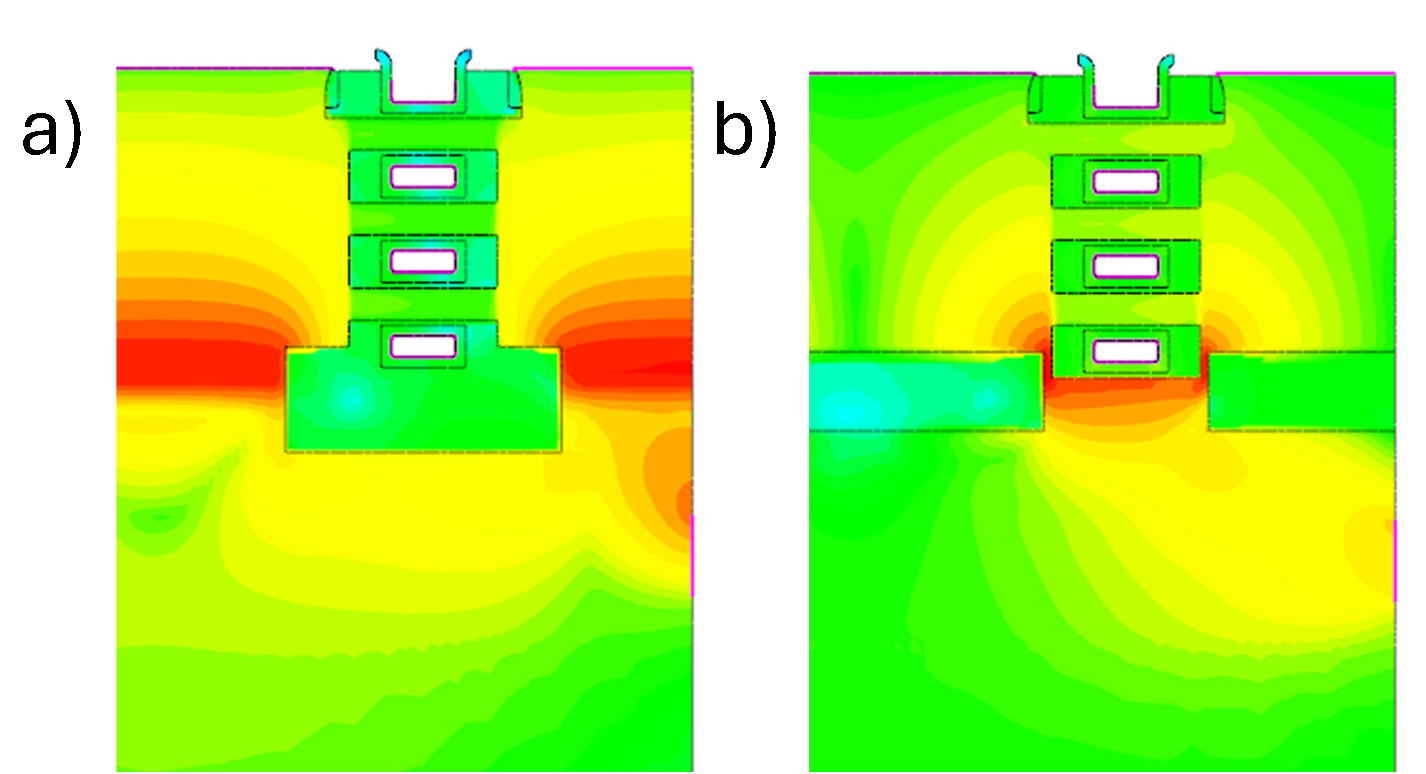}
    \caption{Current leakage for a) C-BDI and b) SD-BDI with underlap. Current density is shown when $V_D=0.8V$ and $V_G=0V$. The substrate side contact is also shown (pink line on the right) and can be a leakage path.}
    \label{fig:CurrentPath}
\end{figure} 

\begin{figure}[b]
    \centering
    \includegraphics[width=1\linewidth]{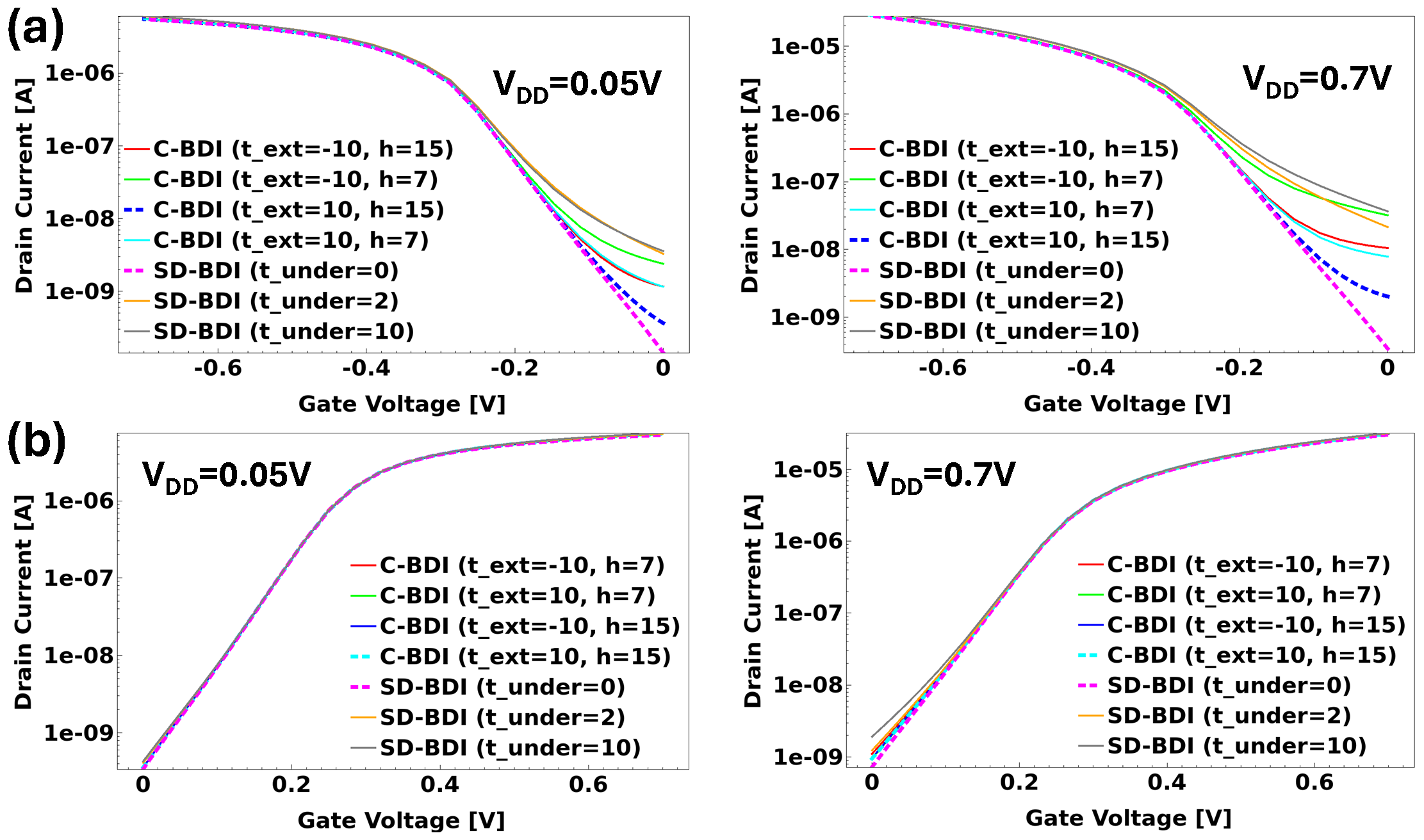}
    \caption{$I_D-V_G$ curves for a) NMOS and b) PMOS GAA with SD-BDI and C-BDI. Legend units are in nm. SD-BDI without underlap (perfect isolation) and C-BDI with the lowest leakage are indicated with dotted curves.}
    \label{fig:IVCurves}
\end{figure}

\begin{figure}[t]  
    \centering
    \includegraphics[width=1\linewidth]{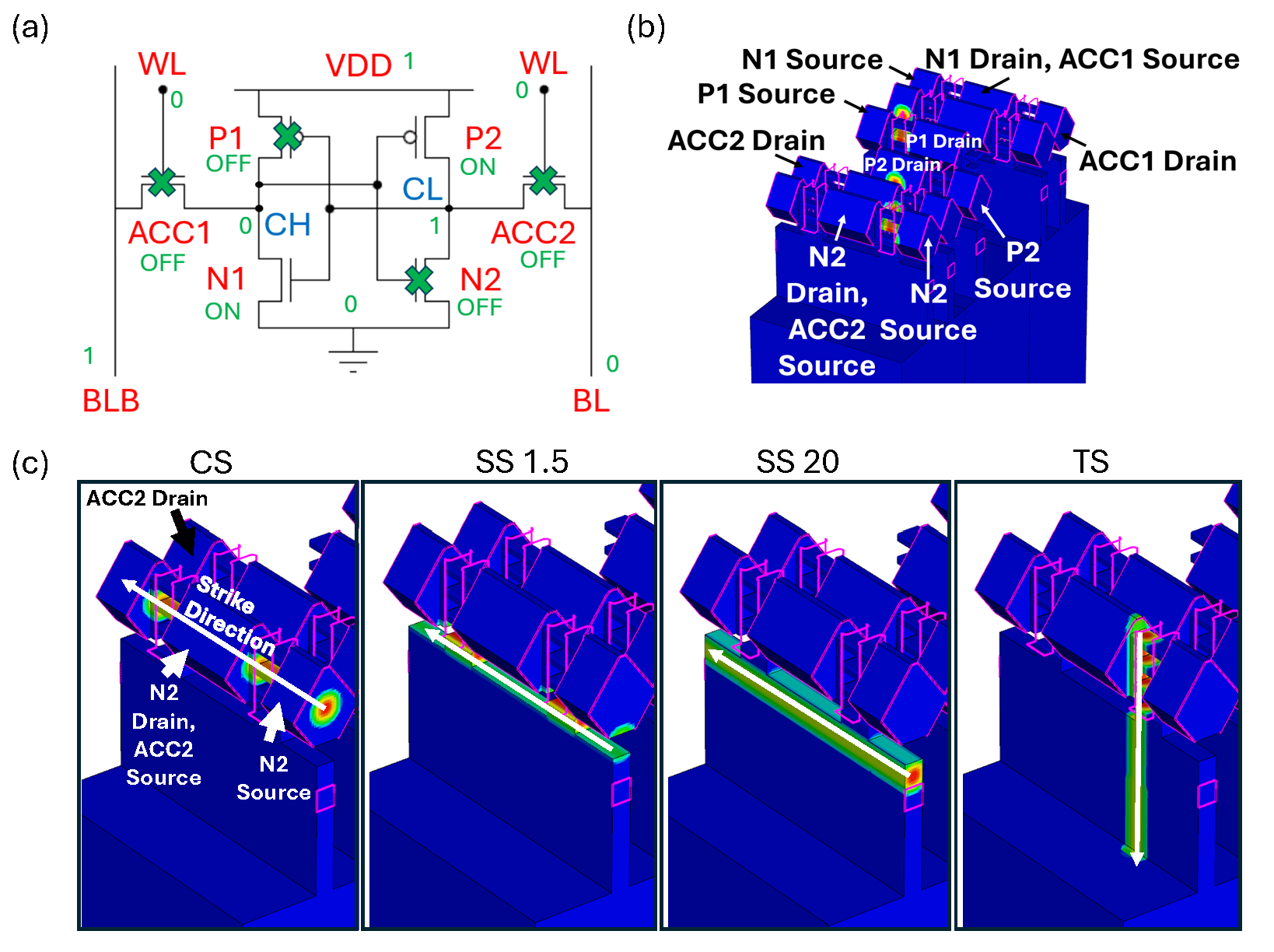}
    \caption{SRAM circuit under the most vulnerable condition (a) and
TCAD structure (b). (b) also shows the diagonal strike-through of
the off-state transistors P1 and N2. c) shows channel strike through
N2 and ACC2 (CS), substrate strike 1.5nm below the epi (SS 1.5),
substrate strike 20nm below the epi (SS 20), and top strike 7nm
from the drain edge (TS).}
    \label{fig:SRAM_setup_figure}
\end{figure} 
    
Leakage can be an issue if the S/D is connected to the substrate, as depicted by the current leakage paths (\textcolor{lightblue}{Fig. \ref{fig:CurrentPath}}). To resolve this issue, both SD-BDI and C-BDI can be combined with PTS. However, as discussed previously, PTS has various disadvantages. Thus, it is beneficial to study devices without PTS. I$_D$V$_G$ curves for SD-BDI and C-BDI (both without PTS) show that the novel C-BDI with $t_{ext} = 10$ nm and $h = 15$ nm has a leakage current only $\sim 3$ times for NMOS (\textcolor{lightblue}{Fig. \ref{fig:IVCurves}a}) and $\sim 5$ times for PMOS (\textcolor{lightblue}{Fig. \ref{fig:IVCurves}b}) compared to the SD-BDI case, even in saturation mode (V$_{D}$=0.7 V). Therefore, the novel C-BDI provides an excellent trade-off between substrate isolation and S/D stress retention. C-BDI also has better substrate isolation compared to SD-BDI with underlaps (\textcolor{lightblue}{Fig. \ref{fig:CurrentPath}}).  

An SRAM is then created to study both self-heating and radiation hardness. \textcolor{lightblue}{Fig. \ref{fig:SRAM_setup_figure}} shows the circuit (top-left) and the full SRAM structure (top-right). ACC1 and ACC2 are the access transistors. N1 and N2 are the pull-down transistors. P1 and P2 are pull-up transistors. BL and BLB are the bit lines. WL is the word line. CH and CL are the storage nodes of the cell.


\subsection{Self-Heating Studies}
\begin{figure}[t] 
    \centering
    \includegraphics[width=1\linewidth]{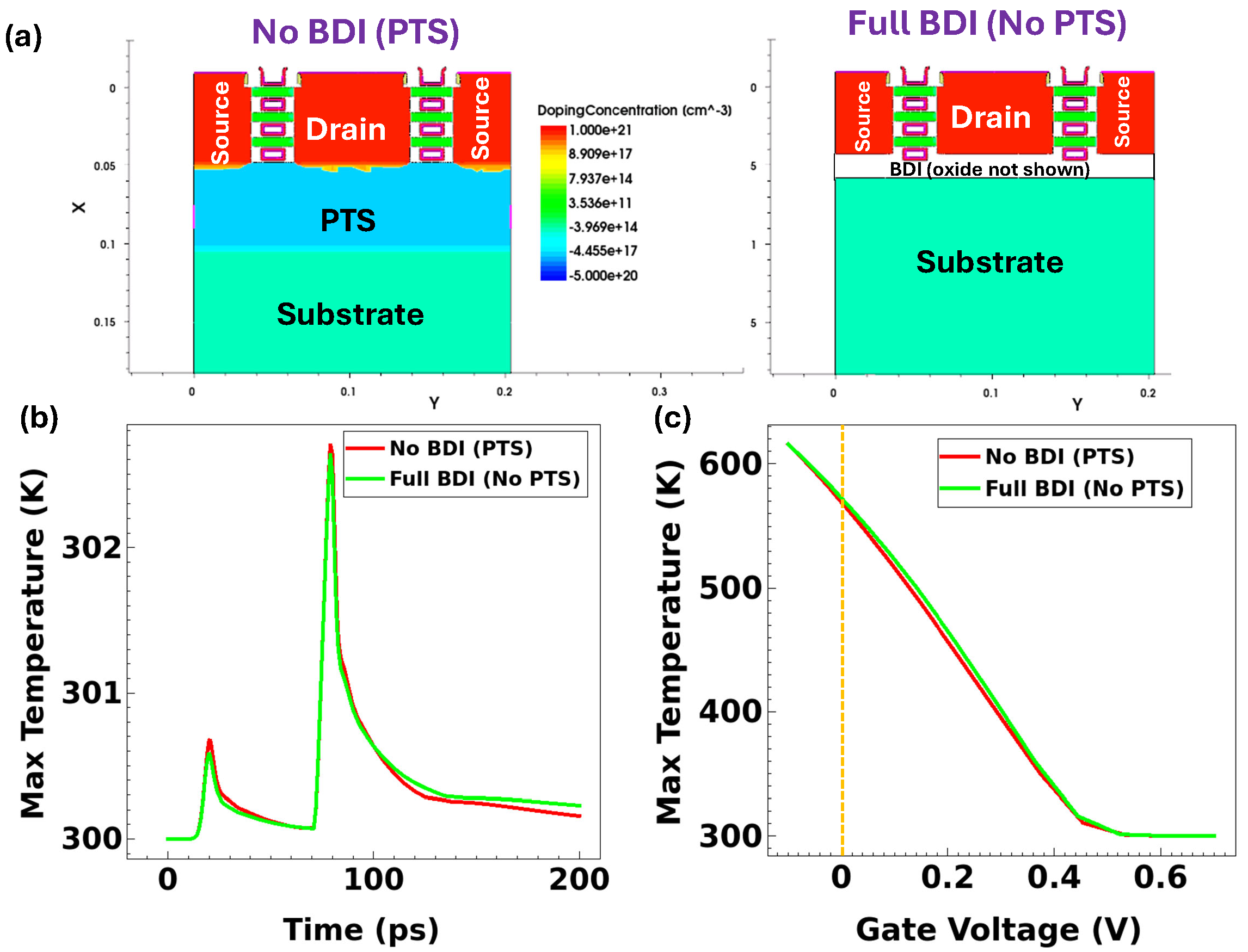}
    \caption{Self-heating simulations for an SRAM. (a) No BDI (PTS is used to avoid leakage) and full-BDI. (b) Self-heating transient simulation showing maximum temperature during a write operation to be at $\sim 80-90ps$. (c) Quasistationary simulation where one PMOS device has $V_{S}$ = 0.7 V, $V_{D}$ = 0 V and its $V_{G}$ is swept with all other transistors off. Full-BDI has a similar self-heating effect as without BDI.}
    \label{fig:SelfHeating}
\end{figure}

Previous studies showed that self-heating is substantially increased due to BDI \cite{b7}. However, those studies used an effective substrate thickness of about 1 $\mu$m (1 $\mu$m Si and thermal resistivity $\theta$ of $5\times 10^{-4}$ K$\cdot$cm$^2$/W). In this study, an SRAM with a full-BDI structure is compared to another SRAM structure without BDI (but with PTS) and with an effective $\sim 500\mu$m Si substrate (0.6 $\mu$m Si and $\theta = 10^{-2}$ K$\cdot$cm$^2$/W) (\textcolor{lightblue}{Fig. \ref{fig:SelfHeating}a}). 

The self-heating is studied using two methods. First, a bit-flip write operation on an SRAM cell is performed because it represents a typical use case. \textcolor{lightblue}{Fig. \ref{fig:SelfHeating}b} shows the maximum temperature through time where only one bit flip is simulated. With and without BDI have similar self-heating. Next, representing a worst-case scenario, a quasistationary simulation is performed where one of the PMOS is biased at $V_{SD}=0.7V$ and $V_G$ is swept. \textcolor{lightblue}{Fig. \ref{fig:SelfHeating}c} shows that BDI is found to have a negligible effect on self-heating. Therefore, in contrast to the previous study, worsening in self-heating due to BDI is negligible even when it is full-BDI, if a realistic effective Si substrate thickness is used. 

\begin{figure}[b] 
    \centering
    \includegraphics[width=1\linewidth]{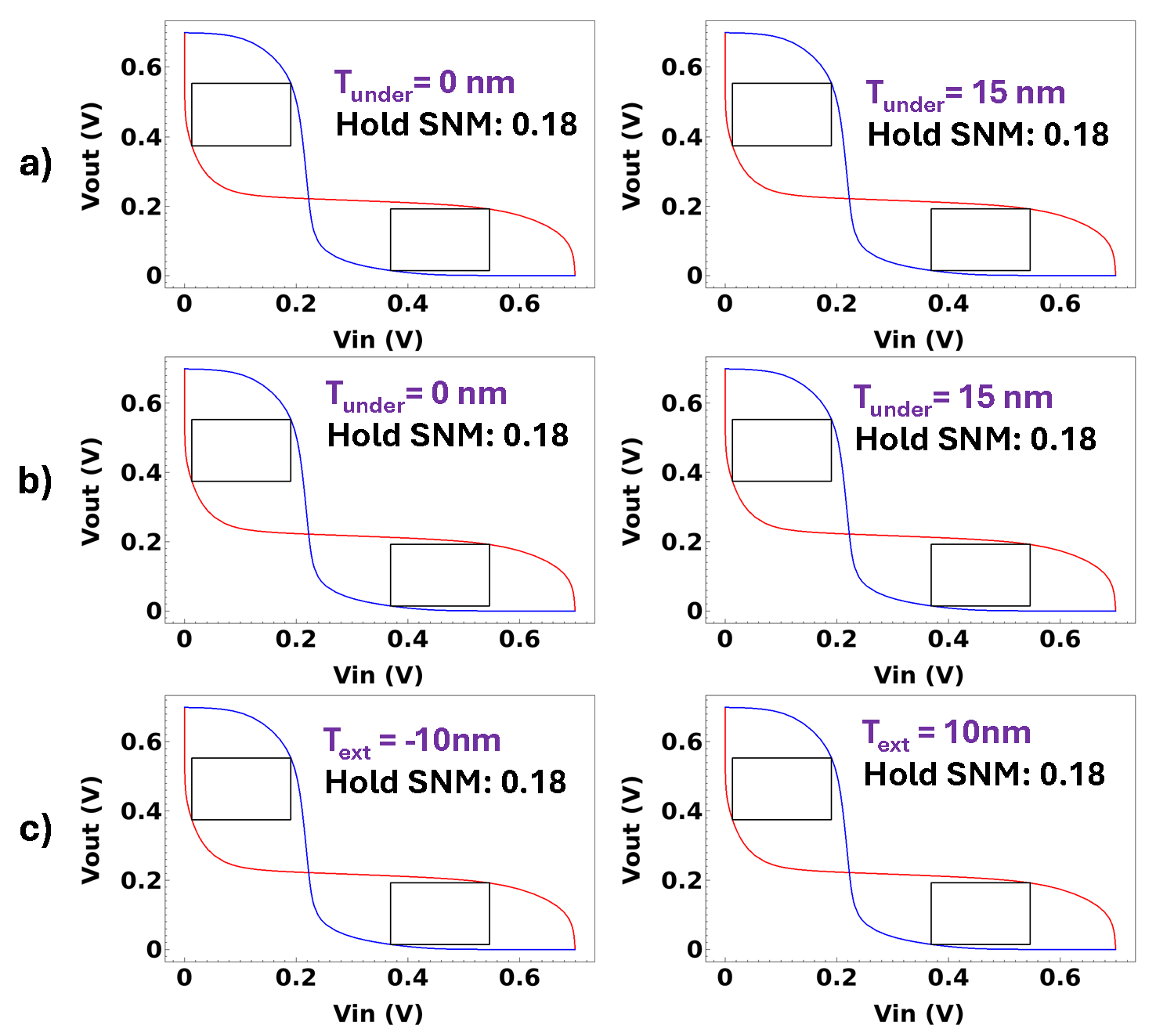}
    \caption{Hold butterfly curves for a) SD-BDI with PTS, b) SD-BDI without PTS and c) C-BDI without PTS at a BDI height of h=7nm. Off-state leakage difference does not affect the SNM.}
    \label{fig:butterfly}
\end{figure}

\begin{figure}[t] 
    \centering
    \includegraphics[width=1\linewidth]{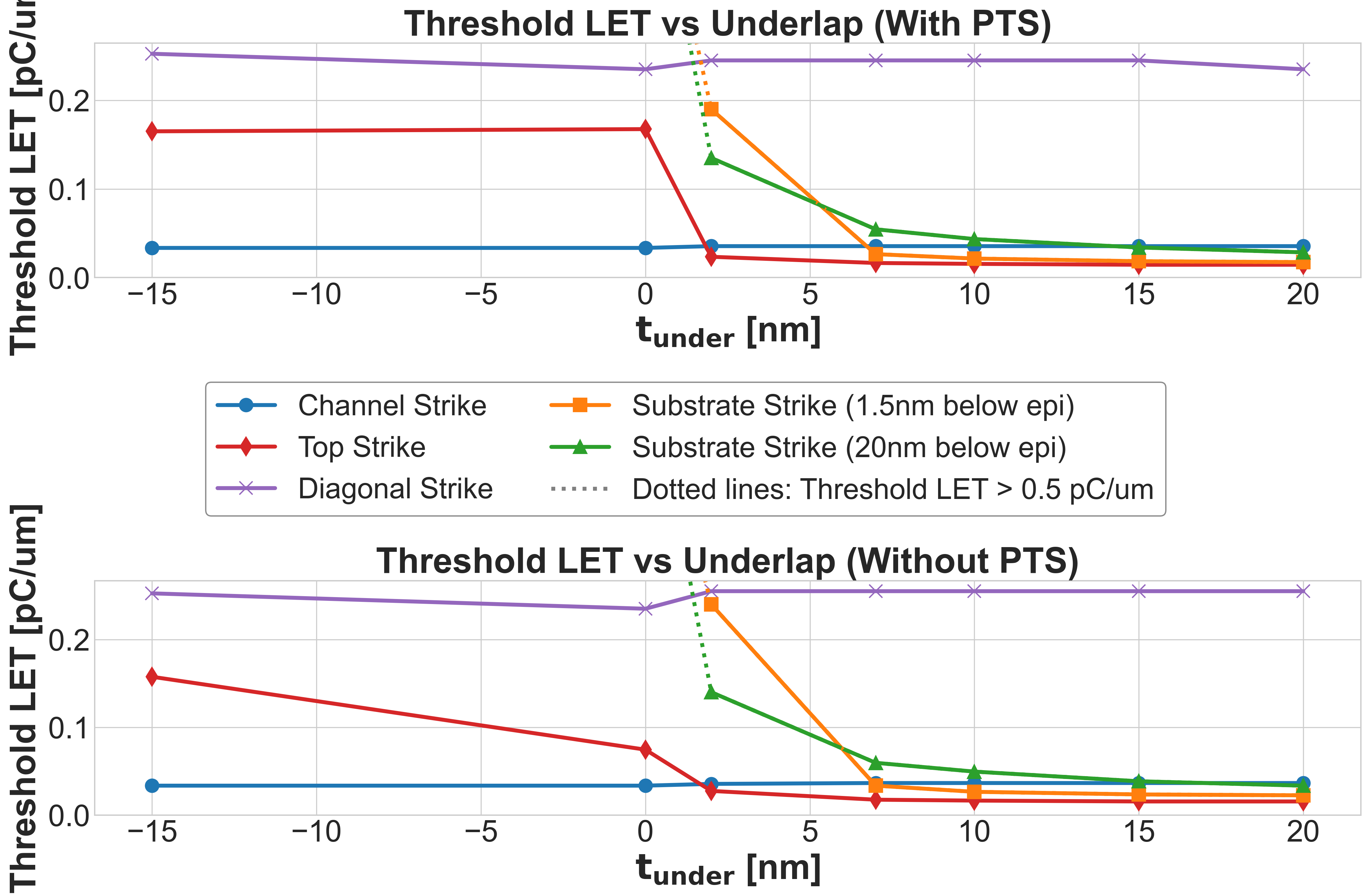}
    \caption{Flipping threshold LET of each radiation strike location for GAA with SD-BDI for h=15nm (top: with PTS; bottom: without PTS). The top strike is the most vulnerable. In some cases, the cell does not flip even with LET $> 0.5$ pC/um (the trend is indicated as dashed lines). This happens for substrate strike when there is no underlap.}
    \label{fig:SDBDIThresholdLET}
\end{figure}

\subsection{Radiation Hardness Studies}
The radiation hardness of GAA-FET SRAMs based on SD-BDI and C-BDI is then studied. For a fair comparison, it is important to ensure the butterfly curves and SNM are similar. SD-BDI and C-BDI with different parameters were shown to have leakage when used without PTS due to the opening to the substrate (\textcolor{lightblue}{Fig. \ref{fig:IVCurves}}). However, leakage current has a small impact on SNM and \textcolor{lightblue}{Fig. \ref{fig:butterfly}} shows that the hold butterfly curves and the SNM of SD-BDI and C-BDI are similar. 

The SRAMs are then initialized into the most vulnerable hold condition shown in \textcolor{lightblue}{Fig. \ref{fig:SRAM_setup_figure}a}. This is when the SRAM is in a hold state while another SRAM in the same column is being written. A single particle can strike multiple off-state transistors at the same time (two NMOS in channel strikes or one NMOS and one PMOS in diagonal strikes). \textcolor{lightblue}{Fig. \ref{fig:SRAM_setup_figure}} shows the 4 key striking locations studied in this work. These are some of the most vulnerable locations as discussed in \cite{b6}\cite{b12}. Alpha particles with different LETs are then varied to determine when the SRAM bit cell will flip for each case. The strikes are modeled as HeavyIon with a Gaussian distribution (peaks at $t=50$ ps and has a standard deviation of 2 ps).  

\begin{figure}[b]
    \centering
    \includegraphics[width=1\linewidth]{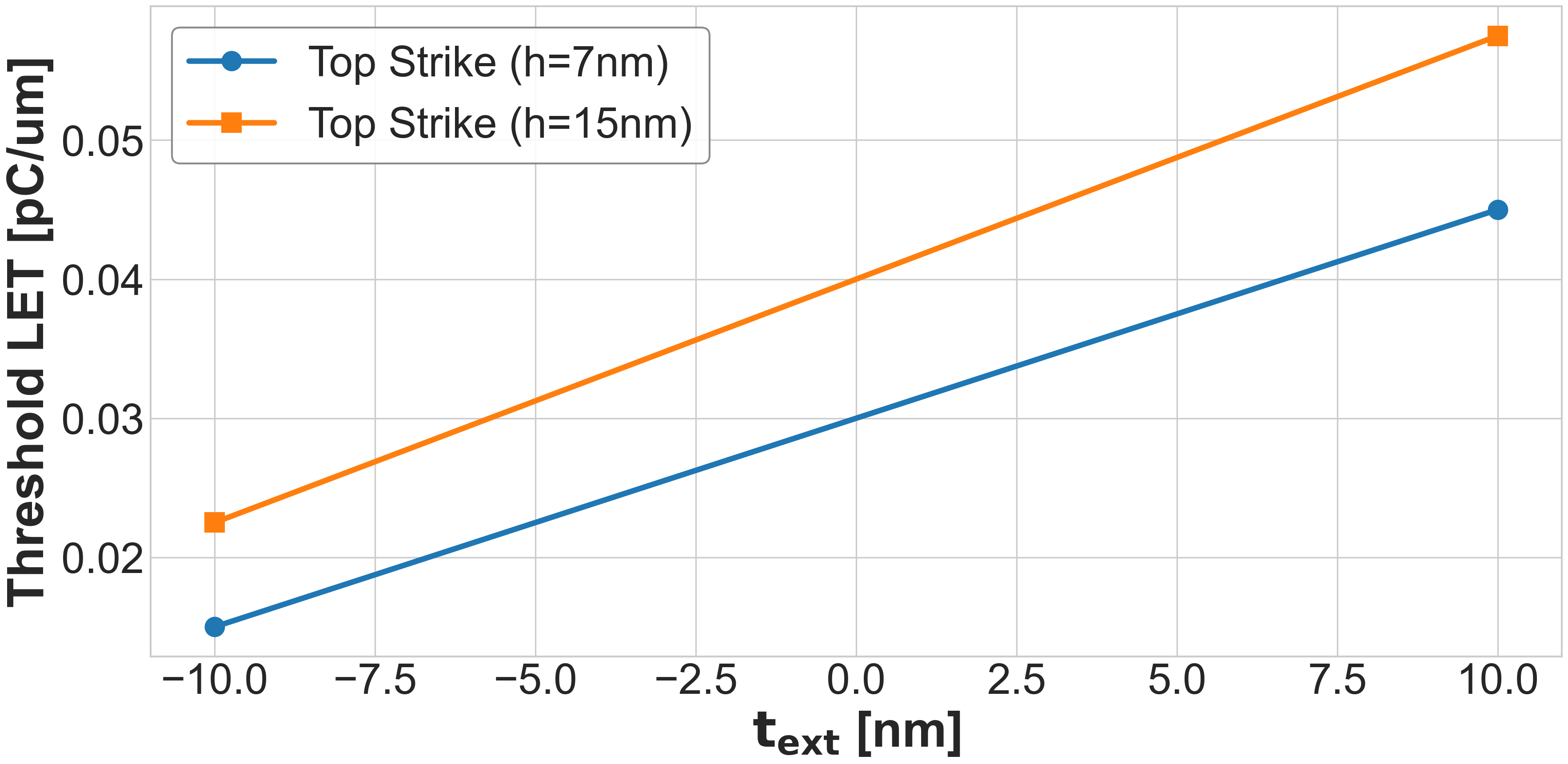}
    \caption{Threshold LET of top strikes for GAA with C-BDI with different $h$ and $t_{ext}$.}
    \label{fig:CBDIThresholdLET}
\end{figure} 

\begin{figure}[t]  
    \centering
    \includegraphics[width=0.95\linewidth]{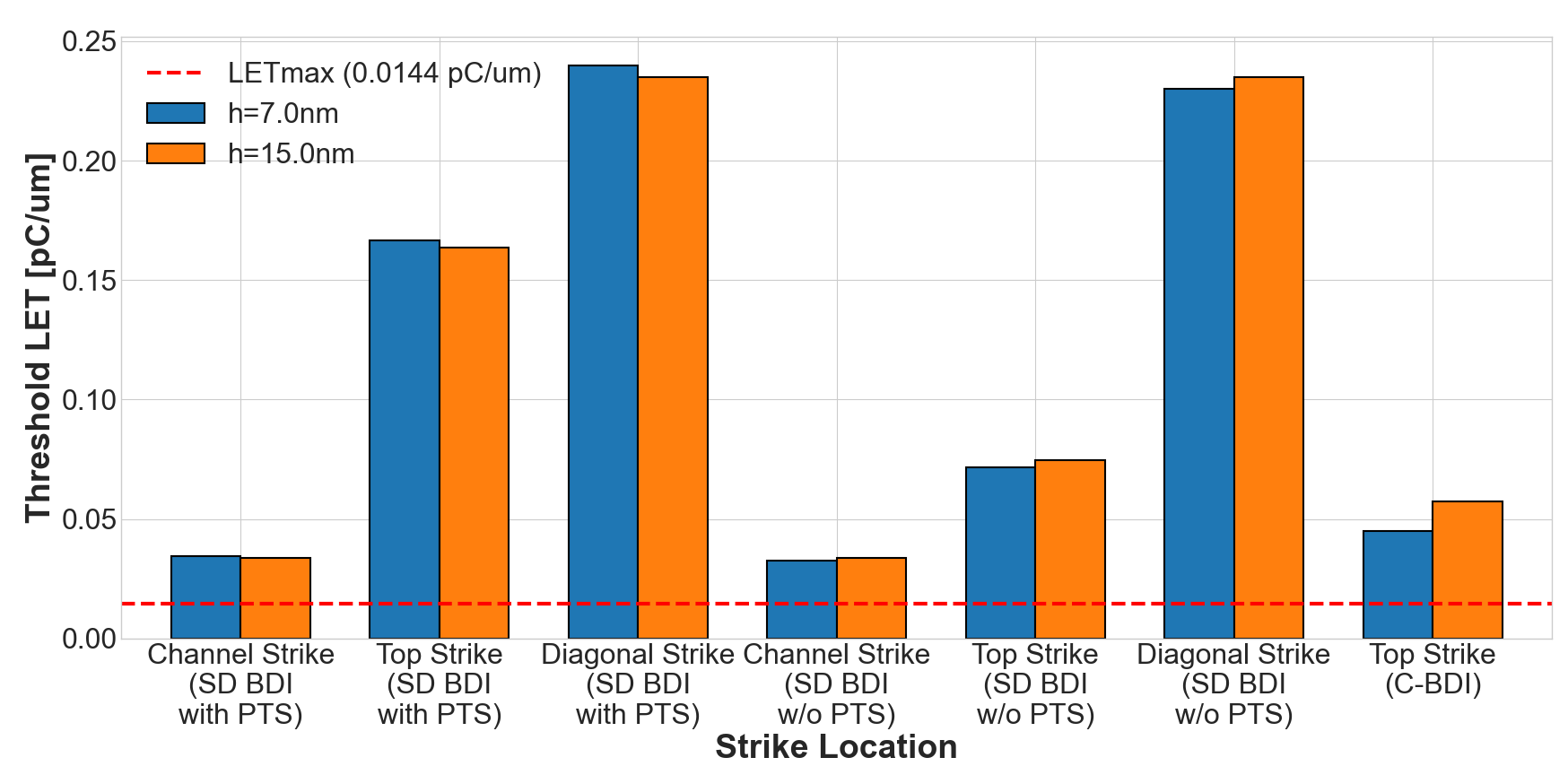}
    \caption{Threshold LET for GAA with varying BDI heights for SD-BDI ($t_{under}=0$) and C-BDI ($t_{ext}=10 nm$). Substrate strikes for SD-BDI are not shown since the SRAM did not flip even at LET greater than 0.5pC/um. Regardless of BDI height, the threshold LET is still greater than 0.0144pC/um (max LET due to an alpha particle). C-BDI has a larger dependence on height.}
    \label{fig:HeightsThreshold}
\end{figure} 

The threshold LET (LET when the SRAM bit cell flips) of each radiation strike location for SD-BDI with/without PTS shows that structures with and without PTS have similar results despite structures without PTS having a large leakage current when $t_{under}$ is positive (\textcolor{lightblue}{Fig. \ref{fig:SDBDIThresholdLET}}). The channel strike and diagonal strike are relatively unaffected by SD-BDI underlap due to their distance away from the BDI layer. In contrast, the substrate strikes and top strike are affected significantly with a positive BDI underlap. When there is full BDI (negative underlap), channel strike is the most vulnerable. As the underlap increases, the radiation hardness for substrate and top strikes decreases nonlinearly, and the top strike becomes the most vulnerable. 

\textcolor{lightblue}{Fig. \ref{fig:CBDIThresholdLET}} shows that the threshold LET of C-BDI of the top strike increases for positive extension and thicker BDI. This is likely because increasing the positive extension and making the BDI thicker increases the oxide area, which would decrease the volume where charge is generated and collected. Less charge allows for a larger threshold LET.  

\textcolor{lightblue}{Fig. \ref{fig:HeightsThreshold}} shows that for all radiation strikes tested, SD-BDI with an underlap of 0 nm and C-BDI with extension of 10 nm, with or without PTS, are all resistant to LET = 0.0144 pC/um, which is the maximum LET value predicted for a single alpha particle strike \cite{b6}. This shows that even with BDI not fully covering the S/D (in the C-BDI case), the radiation hardness is still sufficient. Moreover, \textcolor{lightblue}{Fig. \ref{fig:HeightsThreshold}} shows that BDI thickness has little effect on radiation hardness. 

\section{Conclusion} 
In this work, full-domain 3D TCAD simulations was used to evaluate the performance of GAA-FET SRAMs under various BDI and PTS configurations. We demonstrated that when evaluated with a realistic effective substrate thickness, self-heating from BDI is negligible. We also proposed a novel C-BDI architecture that effectively preserves the connection from S/D and substrate, which mitigates stress relaxation. This design maintains minimal off-state leakage while eliminating the need for a PTS or a BDI that fully isolates the S/D from the substrate. Both SD-BDI and C-BDI are found to improve the overall radiation hardness. Even without fully covering the S/D, they are still predicted to be immune to single alpha particle strikes. 

\section{Acknowledgment}
Sandia National Laboratories is a multimission laboratory managed and operated by National Technology Engineering Solutions of Sandia, LLC, a wholly owned subsidiary of Honeywell International Inc., for the U.S. Department of Energy’s National Nuclear Security Administration under contract DE-NA0003525.

\end{document}